\documentclass[doublecol]{epl2} 

\usepackage{amsmath,amsfonts,amssymb,array,appendix}
\usepackage{shapepar,graphicx,hyperref}
\def\p{\partial}

\def\bn{\textbf{n}}

\def\z{{\tilde{z}}}

\def\const{\rm const}

\title{Instability patterns in ultrathin nematic films:\\ comparison between theory and experiment}

\author{O. V. Manyuhina  \and A.-M. Cazabat  \and M. Ben Amar} 
\institute{Laboratoire de Physique Statistique, Ecole Normale Superieure, UPMC  
Univ Paris 06, Universit\'e Paris Diderot, CNRS, 24 rue Lhomond, 75005  
Paris,  France}

\abstract{Motivated by recent experimental observations~[U. Delabre {\it et al}, Langmuir~{\bf 24}, 3998, 2008] we reconsider an instability of ultrathin nematic films, spread on liquid substrates. Within a continuum elastic theory of liquid crystals, in the harmonic approximation,  we find an analytical expressions for the critical thickness as well as for the critical wavenumber, characterizing the onset of instability towards the stripe phase. Comparing theoretical predictions with experimental observations, we establish the utility of surface-like term such as an azimuthal anchoring.}  

\begin{document}

\maketitle

\section{Introduction}

\begin{figure}
\centering
\includegraphics[width=0.75\linewidth]{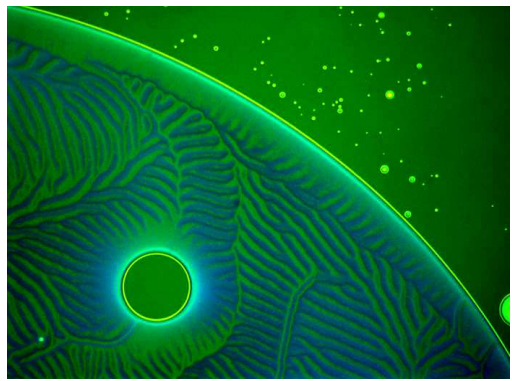}
\caption{Polarised-light microscopy image $1.8\times1.3$~mm of periodically distorted 6CB/water LC at $27^\circ$C. The  thickness of nematic film is $h\approx 0.2~\mu$m and the periodicity of stripes, the distance between two stripes, is $L\approx40~\mu$m$\gg h$. The alternation of colour (dark and light) is associated with violent reorientation ($\sim \pi/2$) of the director $\bn$ in the plane of the image.} 
\label{fig:stripe}
\end{figure}

The formation of spatially periodic patterns in thin nematic films, spread on liquid substrates, is one of the well-known phenomena in physics of liquid crystals~\cite{LP:1990,sparav:1991,LP:1994,sparav:1994,LP:1995,ulysse:2008,ulysse:thesis}. Nematic liquid crystals (LC) are described by the director $\bn$, a unit vector, characterising an averaged preferred orientation of the molecules, which can vary throughout the sample~\cite{degennes:book}. The liquid substrate tends to align this director $\bn$ in the plane of the sample (planar anchoring), whereas the air interface imposes another preferred orientation, the one perpendicular to the interface (homeotropic anchoring). In thick nematic films ($h\gtrsim1\mu m$), the director $\bn$ reorients  in the vertical plane  of the sample in order to relax antagonistic boundary conditions. However, in nematic films with submicron thickness,  the competing interfaces start to ``feel each other'' resulting in undulations of the director. The in-plane symmetry breaking is manifested in the formation of periodic patterns, like  stripes, chevrons, zig-zags, squares~\cite{LP:1994,sparav:1994,ulysse:2008,ulysse:thesis}. An example of stripes  in 6CB nematic film spread on water is shown in Fig.~\ref{fig:stripe}. The wavelength of the observed stripes $2\lesssim L\lesssim 200~\mu$m is typically two orders of magnitude larger compared to the thickness of nematic film $20~\mbox{nm} \lesssim h\lesssim 0.5~\mu$m~\cite{ulysse:2008,ulysse:thesis}. The presence of the free interface and small ratio of $h/L\propto 0.01$ distinguishes our problem from the extensively studied Freedericksz transitions, where liquid crystals form periodic domains in presence of electric and magnetic fields~\cite{williams:1963,lonberg:1985}. Moreover, our system is unique, since the wavelength $L$ of stripes is controlled by the thickness $h$ of nematic film in a non-trivial way. Therefore, one may think of the thickness ``playing the role of the magnetic/electric field'' in the Freedericksz transitions.

First experimental observations of long-wavelength periodic director distortions in thin nematic films were reported in~1990 by Lavrentovich and Pergameshchik~\cite{LP:1990} for 5CB/glycerol LC system. Later, the authors proposed a model~\cite{pergam:1993,LP:1994}, applied to describe experimental data of stripe domains with characteristic wavelength $5<L<200~\mu$m as a function of the film thickness  $0.14<h<0.49~\mu$m. A reasonable agreement between theory and experiment is achieved only by accounting for the splay-bend surface-like elastic term, with the value of elastic constant $K_{13}=-0.205 K_{11}$~\cite{LP:1994},  where $K_{11}=6.2\cdot 10^{-12}$~N~\cite{stewart:book}. Nevertheless, the predicted value for the lower threshold $h_{c1}=0.14~\mu$m is almost one order of magnitude higher than $h_{c1}=20$~nm reported in~\cite{sparav:1994} for 5CB/glycerol and in~\cite{ulysse:2008,ulysse:thesis} for different LC systems. Although the present studies~\cite{LP:1994,sparav:1994,ulysse:2008} agree on the persistence of the striped phase up to the thickness of $h_{c2}\approx 0.5~\mu$m (upper threshold), value for the lower threshold, $h_{c1}$, as well as the mechanism governing the instability  remains an open question. The proposed theoretical models~\cite{sparav:1991,sparav:1994,sparav:1995,pergam:1993,LP:1994} suggest several mechanisms for the formation of stripes, known as $K_{22}$, $K_{24}$ or $K_{24}$-$K_{13}$ mechanisms. These mechanisms play a role only when the anchoring energy is weak and does not exceed the value of $10^{-6}$~J/m$^2$ as was suggested in~\cite{barbero:another,barbero:epl2003}. If we know the elastic constants and the anchoring energies by experimental measurements then we can choose the most suitable model to predict the formation stripes. However, the problem was addressed mostly numerically, assuming  {\it a priori} either idealised values of elastic constants, e.g. $K_{24}=-K_{22}$~\cite{sparav:1991} and $K_{22}=K_{11}$~\cite{sparav:1994,sparav:1995}, or the hybrid aligned cell (HAN) as only possible ground state~\cite{pergam:1993,LP:1994}, which does not allow to understand the physical picture as a whole.

Recent experiments with different LC systems, such as 5CB/glycerol, 6CB/glycerol, 6CB/water, MBBA/glycerol~\cite{ulysse:2008}, suggest that  stripes occur for films as thin as $h_{c1}\simeq20$~nm, with the characteristic wavelength $L\simeq2~\mu$m. To quantify these observations we aim at reconsidering the problem by using an analytical framework, different from  the one developed in~\cite{LP:1990,sparav:1991,LP:1994,LP:1995,sparav:1994} and identifying the critical thickness as well as the critical wavenumber, which determine the onset of the instability from a planar state towards a stripe state. 
Our paper concerns the systematic study of static  instabilities within the Frank--Oseen continuum model of liquid crystals, assuming $K_{33}=K_{11}$~\cite{degennes:book} together with $K_{13}=0$~\cite{yokoyama:1997}, and supplemented by polar and azimuthal anchoring energies~\cite{degennes:book}. Dynamic instabilities  have been previously studied in a more restricted context in~\cite{benamar:2001,cummings:2004}. Our model is a first step for future generalisations.

The influence of the azimuthal anchoring on the threshold thickness was studied in the pioneering work of Sparavigna {\it et al}~\cite{sparav:1991}. The authors have shown for the first time that decreasing the azimuthal anchoring and/or the value of the twist elastic constant $K_{22}$ favours the formation of stripes. However, in that paper the role of the saddle-splay elastic constant $K_{24}$ was disregarded, and only the case $K_{24}=-K_{22}$ was considered. Taking into account both the azimuthal anchoring and the saddle-splay elastic constant would allow to compare theoretical results with experimental findings and to understand the picture as a whole.  In the present  paper we show that azimuthal anchoring, even being vanishingly small, restricts the undulations of the director and breaks the in-plane symmetry. As a result, at the critical point we find a well-defined wavelength for the stripes, which otherwise is infinite. Identifying a finite critical wavenumber and the critical thickness at the instability threshold makes possible a substantial comparison between theory and experimental data~\cite{ulysse:2008,ulysse:thesis}. Let us present first the model.


\section{Theoretical framework}

The nematic liquid crystals are described by an elastic free energy, quadratic in the director derivatives, given by~\cite{degennes:book} 
\begin{multline}\label{eq:Fel}
{\cal F}_{el} = \frac 12\int dV \big\{K_{11}(\nabla, \bn)^2 +K_{22}(\bn,\nabla\times\bn)^2+\\ K_{33}|\bn\!\times\!\nabla\!\times\!\bn|^2
 - (K_{24}\!+\!K_{22})\nabla[\bn(\nabla,\bn) \!-\!(\bn,\nabla)\bn ]\},
\end{multline}
where $K_{11}$, $K_{22}$, $K_{33}$ and $K_{24}$ are elastic moduli, characterising splay, twist, bend, and saddle-splay, respectively. Being a divergence, the last saddle-splay term can be transformed to a surface integral. Although being neglected for thick samples or samples with a strong anchoring boundary conditions, the surface-like terms are crucial in understanding the cause of instabilities in thin nematic films~\cite{sparav:1994,LP:1994,pergam:2000,barbero:2002}. The Ericksen inequalities~\cite{ericksen:1966}, given by
\begin{equation}\label{eq:ericksen}
K_{ii}\geqslant 0,\quad K_{22}+K_{24}\leqslant 2 K_{11},\quad |K_{24}|\leqslant K_{22},
\end{equation}
establish the relationships between the coefficients of the free energy~(\ref{eq:Fel}), which guarantee the stability of the uniform ground state with the director $\bn=\const$. 
The  anchoring energy is usually described by the  Rapini--Papoular potential~\cite{rapini:1969} and is  written in terms of polar angle $\theta$ and azimuthal angle $\varphi$ as 
\begin{multline}\label{eq:Fa}
{\cal F}_a= \frac{W_{\theta 1}}2 \sin^2(\theta_1-\bar\theta_1)+ \frac{W_{\theta 2}}2 \sin^2(\theta_2-\bar\theta_2) + \\+\frac{W_{\varphi 2}}2 \sin^2\theta_2 \sin^2(\varphi_2-\bar\varphi_2),
\end{multline}
where $W_{\theta 1}$ and $W_{\theta 2}$ are the polar anchoring strengths on the liquid substrate and at the air interface, respectively, $W_{\varphi 2}$ is the azimuthal anchoring strength at the air interface, with $W_{\theta 1}>W_{\theta 2}\gg W_{\varphi 2}$. Moreover, on the liquid substrate  the preferred direction is $\bar \theta_1=\pi/2$ (planar anchoring), whereas at the air interface $\bar\theta_2=0$ (homeotropic anchoring). Without loss of generality, the coordinate system is chosen in such a way that the preferred in-plane orientation at the air interface $\bar\varphi_2=0$. In the cases considered experimentally~\cite{ulysse:2008,sparav:1994,LP:1994}, i.e. 5CB/glycerol, 6CB/glycerol, 6CB/water, MBBA/glycerol, we are dealing with a weak anchoring regime, namely $W_{\theta }\propto 10^{-5}$~J/m$^2$, the anchoring extrapolation length being  $L_{\theta 1}=K_{11}/W_{\theta 1}\simeq 0.35~\mu$m and $L_{\theta 2}=K_{11}/W_{\theta 2}\simeq 0.7~\mu$m. The azimuthal anchoring is usually neglected at both interfaces, because the interfaces are isotropic and homogeneous. However, we believe that in the case of strong undulations of the director orientation (see Fig.~\ref{fig:stripe}) considered here, azimuthal anchoring may become essential. A non-zero contribution to the free energy originates from the tendency of $n$-CB molecules with long carbon tails to align along a common direction, thus penalising the perturbation of the director.


In the following, we use the two constant approximation $K_{11}=K_{33}=K$. 
Then two ground states, yielding the minimum of ${\cal F}_{el}+{\cal F}_a$,  are the undistorted planar state with $\bn=(1,0,0)$ or $\theta=\pi/2$ and distorted hybrid aligned state (HAN) with  $\bn=(\sin\theta(z),0,\cos\theta(z))$, $\theta$ is a polar  angle varying along the thickness of the film $h$ as $\theta(z)=\theta_1+(\theta_2-\theta_1) z/h$. The  normalised free energy density is given by  $\omega (\theta_1,\theta_2)= (\theta_1-\theta_2)^2+ h\cos^2\theta_1/L_1+h\sin^2\theta_2/L_2$. If the quadratic form $\omega(\theta_1, \theta_2)$ is positive definite the planar state ($\theta_1=\theta_2=\pi/2$) is stable with respect to  the HAN ($\theta_1\neq\theta_2$) state. 
The anchoring transition, between these two states happens at the Barbero--Barberi critical thickness~\cite{BB:1983} 
\begin{equation}\label{eq:hc}
h_c=L_{\theta2}-L_{\theta1},
\end{equation}
which is approximately $0.35~\mu{\rm m}$ for 5CB/glycerol LC system. However, if we do not require the director $\bn$ to stay in the $xz$-plane, but allow a configuration corresponding to the minimum of the free energy, other equilibrium structures might occur in response to the change of thickness, adopted (and not imposed) by the system. In the following sections we explore a possibility of the transition between a homogeneous planar state and a stripe phase with periodic distortions of the director $\bn$ in $y$-direction (see Fig.~\ref{fig:stripe}).


\section{Variational problem}


To analyse a relative stability of the planar ground state we consider the first variation of the free energy. The 
perturbation of the director $\bn=(1,0,0)$ can be written as  $\bn'=\bn+\delta\bn=\big(\sin(\pi/2+\psi)\cos\phi,\sin(\pi/2+\psi)\sin\phi,\cos(\pi/2+\psi)\big)$, where $\phi$ and $\psi$ are the small variations of the azimuthal and polar angles, respectively. 
The planar ground state is stable if the difference in the free energy $\Delta{\cal F}={\cal F}\{\bn+\delta \bn\}-{\cal F}\{\bn\}>0$, $\forall~\delta\bn$. If there  exists $\bn'$ critical for $\Delta{\cal F}$, so that $\Delta{\cal F}<0$,  then a distorted state is an equilibrium one~\cite{pergam:1993}. Assuming $\p/\p x=0$, since we are interested only in periodic modulation along the $y$-direction, and working in the harmonic approximation, we find from~(\ref{eq:Fel}), (\ref{eq:Fa}) the variation of the free energy as sum of the bulk $f_B$ and the surface $f_S$ contributions
\begin{align}\label{eq:dFtot}
\Delta{\cal F}&=\frac K{2L}\int_0^L dy\,\Big\{f_S+\int_0^h dz\, f_B\Big\},\\[1ex]
f_B&=(\psi_z')^2+t(\psi_y')^2+(\phi_y')^2+t (\phi_z')^2-2\tau\psi_y'\phi_z,\label{eq:FB}\\[1ex]
f_S&=\frac{\psi^2(0)}{L_{\theta 1}} -\frac{\psi^2(h)}{L_{\theta 2}} + \frac{\phi^2(h)}{L_{\varphi 2}} +\notag\\&\quad+ 
2(1-p)\big(\psi(0)\phi_y'(0)-\psi(h)\phi_y'(h)\big),\label{eq:FS}
\end{align}
where  $t=K_{22}/K,\ \tau=1-t,\ p=(K_{22}+K_{24})/K$, $L_{\varphi2}=K/W_{\varphi 2}$  and $L$ is the period of the stripes in $y$-direction. 

In equilibrium, the first variation of the free energy~(\ref{eq:dFtot})  vanishes,  $\delta(\Delta {\cal F})=0$. Thus, the functions $\psi(y,z)$ and $\phi(y,z)$, extremising $\Delta{\cal F}$, can be found from the Euler--Lagrange equations. Looking for a periodic solution written as $\psi(y,z)=f(z)\sin(qy)$ and $\phi(y,z)=g(z)\cos(qy)$~\cite{pergam:1993}, the Euler--Lagrange equations for $f$ and $g$  are
\begin{subequations}\label{eq:EL}
\begin{align}
f_{\z\z}''-t\chi^2 f +\tau \chi g_\z'&=0,\tag{\ref{eq:EL}a}\\
t g_{\z\z}''-\chi^2 g - \tau \chi f_\z'&=0,\tag{\ref{eq:EL}b}
\end{align}
\end{subequations}
where the dimensionless variables $\z=z/h$ and $\chi=q h$ are introduced. A general solution of these two coupled linear differential equations  of second order (Eqs.~(\ref{eq:EL}))  can be written in the form 
\begin{subequations}\label{eq:fg}
\begin{align}
f(z) &= A_1 \sinh(\chi \z)+A_2\cosh(\chi\z)+\notag\\&\quad+A_3 \z\sinh(\chi \z)+A_4\z\cosh(\chi\z),\tag{\ref{eq:fg}a}\\
 g(z) &= B_1 \sinh(\chi \z)+B_2\cosh(\chi\z)+\notag\\&\quad+B_3 \z\sinh(\chi \z)+B_4\z\cosh(\chi\z),\tag{\ref{eq:fg}b}
\end{align}
\end{subequations}
with coefficients $A_i$, $B_i$ dependent on four integration constants $C_1$, $C_2$, $C_3$, $C_4$, given by the following recursive relationships
\begin{subequations}\label{eq:abc}
\begin{alignat}{3}
A_1&=\frac{(1+t)C_1+\tau\chi C_4}{2t\chi},\quad &A_2&=C_3,\tag{\ref{eq:abc}a}\\
A_3&=\frac{\tau(C_2+\chi C_3)}2,\quad &A_4&=-\frac{\tau(C_1+\chi C_4)}{2t},\tag{\ref{eq:abc}b}\\
B_1&=\frac{(1+t)C_2-\tau\chi C_3}{2\chi},\quad &B_2&=C_4,\tag{\ref{eq:abc}c}\\
B_3&=\frac{\tau(C_1+\chi C_4)}{2t},\quad &B_4&=\frac{\tau(C_2+\chi C_3)}2.\tag{\ref{eq:abc}d}
\end{alignat}
\end{subequations}
Note that for $t=1$ ($\tau=0$), system (\ref{eq:EL}) decouples, yielding the coefficients $A_3=A_4=B_3=B_4=0$.


Since, in the linear approximation, the total free energy~(\ref{eq:dFtot}) is a quadratic form in the distortions $\phi$ and $\psi$,  according to~(\ref{eq:fg}), (\ref{eq:abc}) it is also quadratic in the integration constants $C_i$. Substituting the solution  of the Euler--Lagrange equations in~(\ref{eq:dFtot}), and integrating over $z$ and $y$ coordinates, we can rewrite $\Delta {\cal F}$ in the matrix form as~\cite{barbero:2002} 
\begin{equation}\label{eq:M}
\Delta {\cal F}=\frac12\sum_{i,j}M_{ij}C_iC_j, \qquad i,j=1,2,3,4.
\end{equation}
The coefficients $M_{ij}$ of the matrix ${\cal M}$ can be also defined in a simpler way, using integration by parts over $z$ as follows
\begin{multline}\label{eq:dFC}
\frac{\p \Delta F}{\p C_i} = \sum M_{ij} C_j = 
\int_0^h \bigg[\underbrace{\frac{\p f_B}{\p \varphi} -\frac d{dz}\frac{\p f_B}{\p \varphi'_z}}_{\mbox{\small Euler--Lagrange}}\bigg] \frac{\p \varphi} {\p C_i}\,dz +\\+ \underbrace{\bigg(\frac{\p f_S}{\p \varphi}-\frac{\p f_B}{\p \varphi_z'} \bigg)\bigg|_{z=0}}_{\mbox{\small lower boundary}}\!\!\frac{\p \varphi(0)} {\p C_i} + \underbrace{\bigg(\frac{\p f_S}{\p \varphi}+\frac{\p f_B}{\p \varphi_z'} \bigg)\bigg|_{z=h}}_{\mbox{\small upper boundary}}\!\!\frac{\p \varphi(h)} {\p C_i}. 
\end{multline}
The variable $\varphi$ is used for $f$ and $g$, in order to simplify the form of~(\ref{eq:dFC}). Since the functions $f(z)$ and $g(z)$ both satisfy the Euler--Lagrange equations~(\ref{eq:EL}), the expression under the integral in~(\ref{eq:dFC}) is zero. The coefficients $C_i$ in~(\ref{eq:M}) can be found by extremising  $\Delta {\cal F}$ with respect to $C_i$, namely
${\p \Delta F}/{\p C_i} = \sum M_{ij} C_j =0$,
with the latter equation  being  equivalent to the linear combination of the boundary conditions~\cite{barbero:2002} (two boundary conditions for each function). 
A non-trivial solution of the system~ $\sum_j M_{ij}C_j=0$ exists if and only if the determinant of the matrix ${\cal M}$ vanishes, 
\begin{equation}\label{eq:detM0}
\det {\cal M}=0.
\end{equation}
In the case $\det{\cal M}>0$,  the planar state with $C_i=0$ is stable, 
otherwise, an instability  towards a periodically deformed (striped) state occurs. Within a harmonic approximation there is no need in finding the coefficients $C_i$, because the instability threshold is completely determined by the matrix ${\cal M}$. In the next section we find the solution of the governing equation $\det{\cal M}=0$, which defines an implicit relationship between the thickness $h$ of the film and the wavenumber $\chi$, given that other physical parameters of the LC system are known. The minimum of the curves allows to identify both the critical thickness and the critical wavenumber.

\section{The lower threshold}

\begin{figure}
\centering
\raisebox{45mm}{(a)\kern-5mm}\includegraphics[width=0.9\linewidth]{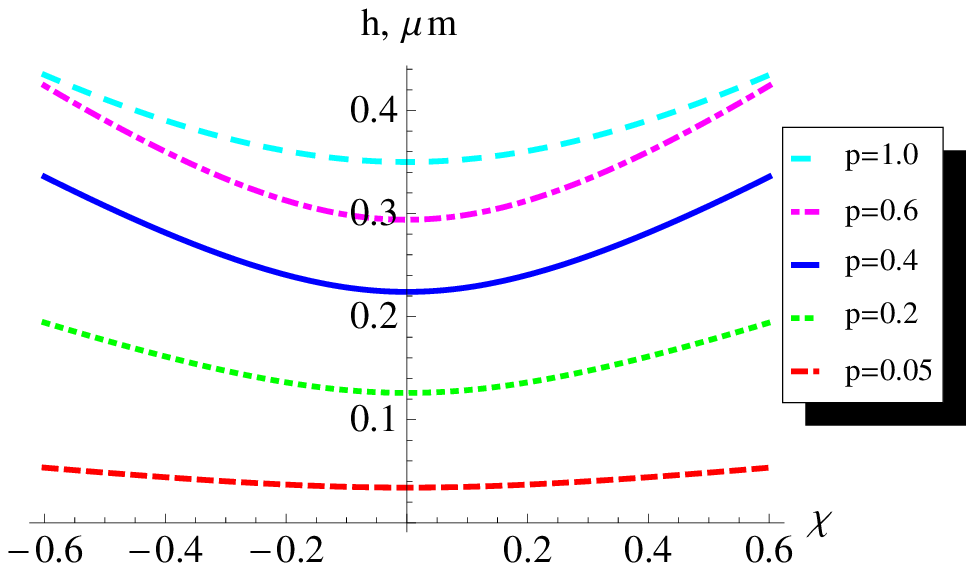}
\vskip3ex
\raisebox{65mm}{(b)\kern2.5cm $p=0.05$\kern-4.5cm}\includegraphics[width=0.9\linewidth]{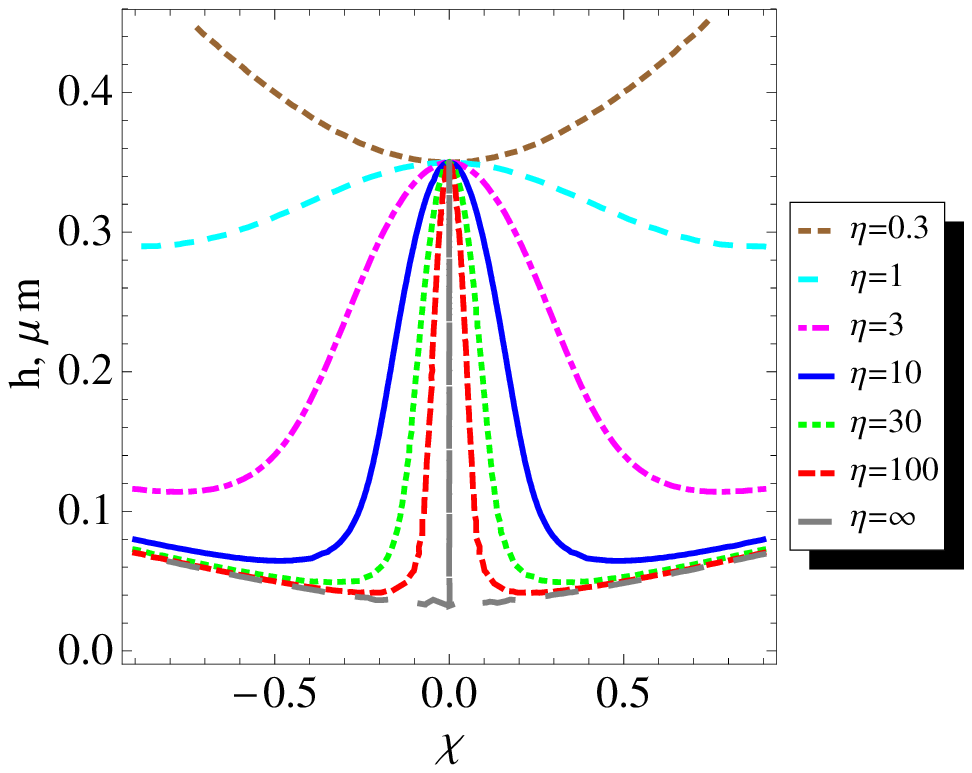}
\caption{The solution of the equation $\det{\cal M}=0$ for the free energy   (a) without azimuthal anchoring~(\ref{eq:M4}),  (b)  with azimuthal anchoring as in~(\ref{eq:M4a}). The typical values $L_{\theta1}=0.35~\mu$m, $L_{\theta2}=0.7~\mu$m and $t=0.6$ are chosen. The curves correspond to different values of $p=t+K_{24}/K$, satisfying Ericksen inequalities~(\ref{eq:ericksen}), and $\eta=L_{\varphi2}/L_{\theta2}$.  (a) The minimum value of $h(\chi)$ defines the critical thickness $h_{c1}$ (\ref{eq:hc1p})  achieved at $\chi=0$, $\forall~p$. (b) Two symmetric minima with non-zero $\chi_{c1}$ and $h_{c1}$ separated by a maximum $h_{c1}=L_{\theta2}-L_{\theta1}=0.35~\mu$m at $\chi=0$ for $L_{\varphi2}>L_{\varphi2}^{\rm cr}\simeq0.455~\mu$m ($\eta>0.65$). In degenerate case of vanishingly small azimuthal anchoring ($L_{\varphi2}\to \infty$) the critical parameters for (a) and (b) at $p=0.05$ coincide.} 
\label{fig:h2x}
\end{figure}

\begin{figure}
\centering
\raisebox{48mm}{(a)}\includegraphics[width=0.86\linewidth]{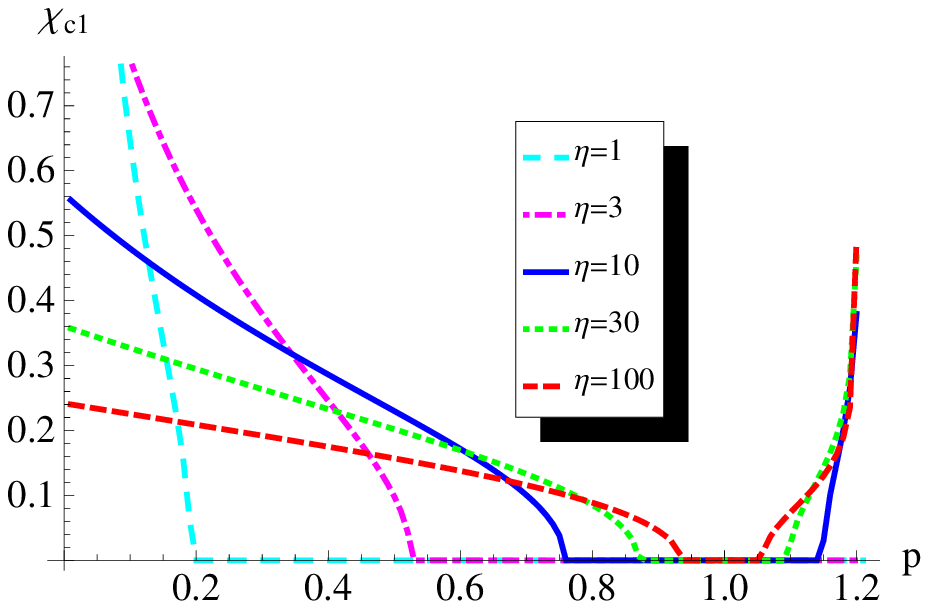}
\hfil
\raisebox{48mm}{(b)}\includegraphics[width=0.86\linewidth]{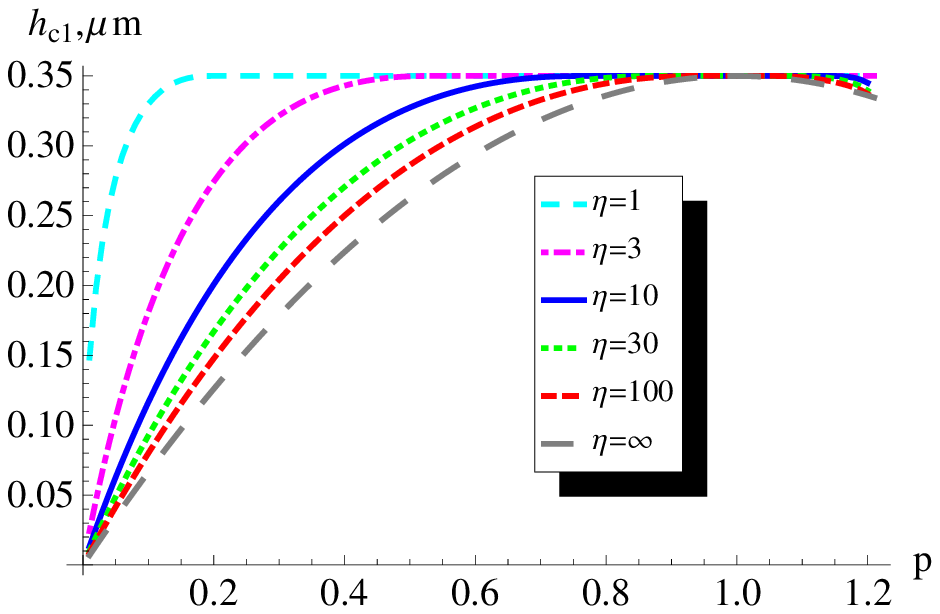}
\caption{Critical threshold $h_{c1}$ and $\chi_{c1}=2\pi h_{c1}/L$  as function of $p=t+K_{24}/K<2t$ calculated for different $\eta=L_{\varphi2}/L_{\theta2}$. The curves correspond to the local minima of $\det{\cal M}$ (\ref{eq:Mexp}).  The values for  $t=0.6$, $L_{\theta 1}=0.35~\mu$m, $L_{\theta 2}=0.7~\mu$m are chosen. The long-dashed line in (b) corresponds to the case of vanishing azimuthal anchoring ($L_{\varphi2}\to\infty$) and is given by (\ref{eq:hc1p}). Strong azimuthal anchoring ($\eta\propto 1$) suppresses the formation of stripes $\chi_{c1}=0$ (a), as predicted in~\cite{sparav:1991}.} 
\label{fig:hcxc}
\end{figure}

We first consider an instability of the planar state towards a stripe phase,  neglecting  the azimuthal anchoring contribution ($W_{\varphi2}=0$) to the surface free energy~(\ref{eq:FS}). 
By substituting the solution of the Euler--Lagrange equations~(\ref{eq:fg}) into~(\ref{eq:FB}), (\ref{eq:FS}), and using~(\ref{eq:dFC}), we find the coefficients $M_{ij}$ and consequently the determinant  of the matrix ${\cal M}$ as 
\begin{multline}\label{eq:M4}
\det {\cal M}\! =\! \frac{(1 - t)^2 \chi^2 - (1 + t)^2 \sinh^2 \chi}{L_{\theta1} L_{\theta2} t^2 \chi ^2} \cdot 
\big\{4 h^2 t^2(\sinh\chi)^2 + \\+  h (L_{\theta1} - L_{\theta2}) p t \chi \big[2p (1 - t)\chi- (p-(4-p)t) \sinh(2 \chi)\big]  +\\  L_{\theta1} L_{\theta2} p^2 \chi^2 \big[p^2 (1 - t)^2 \chi^2 - (p-(4- p)t)^2(\sinh\chi)^2\big]\big\}.
\end{multline}
Figure~\ref{fig:h2x}a shows a stable solution of the equation $\det {\cal M}(h,\chi)=0$  for a typical value of $t=0.6$~\cite{stewart:book,LP:1994} and $L_{\theta 1}=0.35~\mu$m, $L_{\theta 2}=0.7~\mu$m~\cite{sparav:1995,ulysse:thesis}, corresponding to 5CB on glycerol at room temperature. The minimum value of $h$ on these curves yields the critical thickness $h_{c1}$, when the non-zero perturbations of the director $\bn$ may appear. The critical wavenumber $\chi_{c1}$ at $h=h_{c1}$ turns out to be zero irrespective of the value of $p$, meaning that the perturbations have an infinite wavelength. This finding contradicts experimental observations~\cite{ulysse:2008,sparav:1994} of a finite wavelength $L\approx 2~\mu$m at the lower threshold thickness.  Nevertheless, slightly above the threshold ($h>h_{c1}$) the model predicts a finite and small dimensionless wavenumber $\chi$. The expansion of (\ref{eq:M4}) around $\chi_{c1}\equiv 0$ gives $\det{\cal M}=-16h(h-h_{c1})t \chi^2/( L_{\theta 1} L_{\theta 2}) +O(\chi^4)$ with the critical thickness 
\begin{equation}\label{eq:hc1p}
h_{c1}=(L_{\theta 2} - L_{\theta 1})(2-p)p,\qquad 0\leqslant p\leqslant 2t\leqslant 2, 
\end{equation}
where $p=t+K_{24}/K$ satisfies the Ericksen inequalities (\ref{eq:ericksen}). Below $h_{c1}$,  the homogeneous planar state is stable ($\det{\cal M}>0$), whereas above $h_{c1}$ the periodically modulated stripe phase is an equilibrium one ($\det{\cal M}<0$). Let us point out that i) the solution~(\ref{eq:hc1p}) gives the minimum of $h(\chi)$, since the second derivative $\p^2 h/\p\chi^2\propto p(p-2t)/(L_{\theta 1} - L_{\theta 2})+O(\chi^2)>0$, is positive and ii) another formal solution for the critical thickness, $h_{c1}=0$, is stable only for $p>2t$, which violates the Ericksen inequalities. In the experiments on different LC systems~\cite{ulysse:2008}, $h_{c1}=40\pm20$~nm, which is theoretically achievable only for a small values of $p$, when $K_{24}\gtrapprox-K_{22}\simeq -0.6 K$. This fits in the  range of the values  $-0.6K\leqslant K_{24}\leqslant 0.6 K$, measured experimentally for 5CB LC compound~\cite{allender:1991}. However, in the case of an exact equality, namely $K_{24}=\mp K_{22}$ ($p=0$ or $p=2t$), the second derivative $\p^2 h/\p\chi^2$ vanishes, resulting in a saddle-point rather than a minimum for the curves $h(\chi)$. Notice that, according to~(\ref{eq:hc1p}), the critical thickness does not depend on the value of the twist elastic constant~$t$, as was mentioned by Pergameshchik~\cite{pergam:1993} as well. The plot of~(\ref{eq:hc1p}), shown in Fig.~\ref{fig:hcxc}, is in agreement with calculations of Sparavigna {\it et al} in~\cite{sparav:1994}.  In conclusion, the considered model provides a finite value for the critical thickness, however, at the critical point $\chi$ is zero (infinite wavelength), which contradicts experimental observations~\cite{ulysse:2008}. Our result is robust and cannot be improved by taking into account the $K_{13}$ surface-term, which would modify (\ref{eq:dFC}) and would result in the rescaling of the critical thickness, achieved again at  $\chi=0$~\cite{hc1s}.

Now we aim at identifying the critical thickness and the critical wavenumber, taking into account the azimuthal anchoring $W_{\varphi 2}$ on the nematic--air interface. In essence, the expected symmetry breaking should appear in the form of the determinant of the matrix ${\cal M}$, given by
\begin{multline}\label{eq:M4a}
\det {\cal M} = \frac{(1 - t)^2 \chi^2 - (1 + t)^2 \sinh^2 \chi}{L_{\theta1} L_{\theta2}L_{\varphi2} t^2 \chi ^3} 
\cdot \\
\Big\{h^3 t [(1 + t) \sinh(2 \chi) - 2 (1 - t) \chi ] + \\+h^2  \chi \big[ 4  t^2 (L_{\theta1}+L_{\varphi 2}(\sinh\chi)^2 - L_{\theta 2} (\cosh\chi)^2)  + \\+L_{\theta 1} p (p(1 - t)^2 \chi^2 - (1 + t) (p - (4 - p) t)(\sinh\chi)^2) \big] + \\+
h p t \chi^2 \big[2 \big(L_{\theta 1} (L_{\theta 2} + L_{\varphi 2}) - L_{\theta 2} L_{\varphi 2} \big) p (1 - t) \chi + \\+\big(L_{\theta 1} (L_{\theta 2} - L_{\varphi 2}) +  L_{\theta 2} L_{\varphi 2}\big) (p - (4 - p) t) \sinh(2 \chi)\big]  +  \\
L_{\theta 1} L_{\theta 2} L_{\varphi 2} p^2 \chi^3 \big[p^2 (1-t)^2 \chi^2 - (p-(4-p)t)^2 (\sinh\chi)^2\big] 
\Big\}.
\end{multline}
Indeed, $\det{\cal M}=0$ becomes a cubic equation with respect to $h$, rather than a quadratic one when $W_{\varphi2}=0$. The solution of this equation is plotted in Fig.~\ref{fig:h2x}b for different values of the azimuthal anchoring $W_{\varphi2}\simeq 3\cdot10^{-5}\div 10^{-7}$~J/m$^2$ ($L_{\varphi2}\simeq 0.21\div70~\mu$m or $\eta=L_{\varphi2}/L_{\theta2}=0.3\div100$), which is an unknown parameter  for the studied LC systems~\cite{ulysse:2008,ulysse:thesis}. Depending on the value of  $\eta$ the  curves  exhibit either a single minimum at $\{h_{c1},\chi_{c1}\}=\{L_{\theta2}-L_{\theta1},0\}$ or two symmetric minima at $\chi=\pm \chi_{c1}\neq0$. In terms of the  Landau theory of a second-order phase transitions,  one might consider the wavenumber as an order parameter, being zero in homogeneous planar state and  non-zero (or small) in a stripe phase, and the azimuthal anchoring playing the role of control parameter (e.g. temperature)~\cite{chaikin:book}. The solution at $\chi_{c1}=0$ looses its stability when $\p^2 h/\p\chi^2|_{h\to L_{\theta2}-L_{\theta1}}<0$  or 
\begin{multline}\label{eq:Lphicr}
L_{\varphi2}>L_{\varphi2}^{\rm cr}= \frac{1}{3 (L_{\theta1} - L_{\theta2}) (1 - p)^2 t} \cdot\big[L_{\theta2}^2 (1 - 2 t) +\\ L_{\theta1}^2 (1+t - 3 p (1 - p +t)) + L_{\theta1} L_{\theta2} (t-2+3 p (1 - t) )\big].
\end{multline}
In the case $p=0.05$, illustrated in Fig.~\ref{fig:h2x}b, $L_{\varphi2}^{\rm cr}\simeq 0.455~\mu$m. To study the behaviour of the curves around the critical point we expand (\ref{eq:M4a}) in powers of $\chi$ as
\begin{equation}\label{eq:Mexp}
\det{\cal M}= a_0(h) + a_2(h)\chi^2+a_4(h)\chi^4+O(\chi^6),
\end{equation}
where $a_0=- {h^2 (h+ L_{\theta1} - L_{\theta2})}/(L_{\theta1}L_{\theta2}L_{\varphi2})$, the coefficients $a_2$ and $a_4$ contain bulky expressions, which is not necessary to write explicitely. Unless $a_4$ is always positive, the next order term should be added to the series~(\ref{eq:Mexp}). When $a_2>0$ (equivalent to $L_{\varphi2}<L_{\varphi2}^{\rm cr}$~(\ref{eq:Lphicr})), the  equilibrium solution is $\chi_{c1}=0$, and $h_{c1}=L_{\theta2}-L_{\theta1}$ is given by the condition $\det{\cal M}=a_0=0$. For $a_2<0$, two symmetric minima occur at $\chi_{c1}=\pm\sqrt{-a_2/(2a_4)}$, yielding the following condition $\det{\cal M}=a_0-a_2^2/(4a_4)=0$ for the critical thickness $h_{c1}$.  The solution of these equations is shown in Fig.~\ref{fig:hcxc}, assuming the values $L_{\theta i}$ and $t$ for 5CB/glycerol. The behaviour of the curves in Figs.~\ref{fig:h2x} and \ref{fig:hcxc} for other LC systems should be qualitatively the same, because it results from an intrinsic symmetry of the problem and it is not caused by a particular choice of parameters. The plotted equilibrium values for $\chi_{c1}$ and $h_{c1}$ correspond to the local minima of $\det{\cal M}$ (\ref{eq:M4a}), shown in Fig.~\ref{fig:h2x}. The employed expansion of $\det{\cal M}$ is justified only if $\chi\lesssim1$, which holds for small azimuthal anchoring in the whole range of $p$ (see Fig.~\ref{fig:hcxc}). Note that in the limit of the vanishing azimuthal anchoring ($L_{\varphi2}\to \infty$) the solution for the critical thickness $h_{c1}$ converges to~(\ref{eq:hc1p}). The qualitative as well as the quantitative agreement between theory and experiment is achieved for $p\propto 0$ ($K_{24}\propto -K_{22}$) and small, but {\it non-zero}, value of $W_{\varphi2}\propto 10^{-7}$~J/m$^2$.

\section{Concluding remarks}

In this paper we considered  the onset of instability of planar nematic films, subjected to competing boundary conditions,  towards a periodically deformed state (stripes). In the harmonic approximation, we found an exact solution for the variational problem and were able to identify the critical thickness as well as the critical wavenumber, characterising the lower threshold instability. The analysis is performed within the continuum theory framework, which does not necessarily hold for films with thickness of the tens of nanometers. Nevertheless, by taking into account the azimuthal anchoring at the nematic--air interface, we found a reasonable agreement between theory and experiment~\cite{ulysse:2008,ulysse:thesis}. Considering only the saddle-splay surface term, turns out to be insufficient to identify a finite wavelength for stripes at the lower threshold. One could study the described phenomena  above the threshold and decide to analyse the experimental data as in~\cite{LP:1994}. However, this would require a sophisticated non-linear analysis for the free energy. Finding the upper threshold can give additional insight into experimental data and will allow to discuss the possibility for the formation of stripes in thick nematic samples observed in~\cite{barbero:epl2003}. The  study of the instability from a hybrid aligned nematic towards stripe phase will be fulfilled in the future within the same framework.

\acknowledgments The authors gratefully acknowledge fruitful and stimulating discussions with Gaetano Napoli, Ulysse Delabre, C\'eline Richard, and Paolo Galatola. This work was partially supported by the French National Research Agency
(ANR), grant ANR-07-BLAN-0158.


\begin{thebibliography}{10}
\bibitem{hc1s}
In presence of the $K_{13}$ surface-like term the expression for the critical
  thickness~(\ref{eq:hc1p}) is modified as follows $h_{c1}=(L_{\theta
  2}-L_{\theta1})(1-(1-p+s)^2+2s)$, where $s=K_{13}/K$. This result, obtained
  within the presented theoretical framework, will be published elsewhere.
\end{thebibliography}
\end{document}